\documentclass[11pt,singlecolumn,superscriptaddress]{revtex4-1}
\pdfoutput=1

\usepackage{graphicx}
\usepackage{dcolumn}
\usepackage{bm}
\usepackage{color}
\usepackage{epstopdf}
\newcommand{\enquote}[1]{#1}

\begin{document}
\title{Identifying defect-related quantum emitters in monolayer WSe$_2$}

\author{Jianchen Dang}
\affiliation{Beijing National Laboratory for Condensed Matter Physics, Institute of Physics, Chinese Academy of Sciences, Beijing 100190, China}
\affiliation{CAS Center for Excellence in Topological Quantum Computation and School of Physical Sciences, University of Chinese Academy of Sciences, Beijing 100049, China}
\author{Sibai Sun}
\affiliation{Beijing National Laboratory for Condensed Matter Physics, Institute of Physics, Chinese Academy of Sciences, Beijing 100190, China}
\affiliation{CAS Center for Excellence in Topological Quantum Computation and School of Physical Sciences, University of Chinese Academy of Sciences, Beijing 100049, China}

\author{Xin Xie}
\affiliation{Beijing National Laboratory for Condensed Matter Physics, Institute of Physics, Chinese Academy of Sciences, Beijing 100190, China}
\affiliation{CAS Center for Excellence in Topological Quantum Computation and School of Physical Sciences, University of Chinese Academy of Sciences, Beijing 100049, China}
\author{Yang Yu}
\affiliation{Beijing National Laboratory for Condensed Matter Physics, Institute of Physics, Chinese Academy of Sciences, Beijing 100190, China}
\affiliation{CAS Center for Excellence in Topological Quantum Computation and School of Physical Sciences, University of Chinese Academy of Sciences, Beijing 100049, China}

\author{Kai Peng}
\affiliation{Beijing National Laboratory for Condensed Matter Physics, Institute of Physics, Chinese Academy of Sciences, Beijing 100190, China}
\affiliation{CAS Center for Excellence in Topological Quantum Computation and School of Physical Sciences, University of Chinese Academy of Sciences, Beijing 100049, China}
\author{Chenjiang Qian}
\affiliation{Beijing National Laboratory for Condensed Matter Physics, Institute of Physics, Chinese Academy of Sciences, Beijing 100190, China}
\affiliation{CAS Center for Excellence in Topological Quantum Computation and School of Physical Sciences, University of Chinese Academy of Sciences, Beijing 100049, China}
\author{Shiyao Wu}
\affiliation{Beijing National Laboratory for Condensed Matter Physics, Institute of Physics, Chinese Academy of Sciences, Beijing 100190, China}
\affiliation{CAS Center for Excellence in Topological Quantum Computation and School of Physical Sciences, University of Chinese Academy of Sciences, Beijing 100049, China}
\author{Feilong Song}
\affiliation{Beijing National Laboratory for Condensed Matter Physics, Institute of Physics, Chinese Academy of Sciences, Beijing 100190, China}
\affiliation{CAS Center for Excellence in Topological Quantum Computation and School of Physical Sciences, University of Chinese Academy of Sciences, Beijing 100049, China}
\author{Jingnan Yang}
\affiliation{Beijing National Laboratory for Condensed Matter Physics, Institute of Physics, Chinese Academy of Sciences, Beijing 100190, China}
\affiliation{CAS Center for Excellence in Topological Quantum Computation and School of Physical Sciences, University of Chinese Academy of Sciences, Beijing 100049, China}
\author{Shan Xiao}
\affiliation{Beijing National Laboratory for Condensed Matter Physics, Institute of Physics, Chinese Academy of Sciences, Beijing 100190, China}
\affiliation{CAS Center for Excellence in Topological Quantum Computation and School of Physical Sciences, University of Chinese Academy of Sciences, Beijing 100049, China}
\author{Longlong Yang}
\affiliation{Beijing National Laboratory for Condensed Matter Physics, Institute of Physics, Chinese Academy of Sciences, Beijing 100190, China}
\affiliation{CAS Center for Excellence in Topological Quantum Computation and School of Physical Sciences, University of Chinese Academy of Sciences, Beijing 100049, China}

\author{Yunuan Wang}
\affiliation{Beijing National Laboratory for Condensed Matter Physics, Institute of Physics, Chinese Academy of Sciences, Beijing 100190, China}
\affiliation{Key Laboratory of Luminescence and Optical Information, Ministry of Education, Beijing Jiaotong University, Beijing 100044, China}
\author{M. A. Rafiq}
\affiliation{Department of Physics and Applied Mathematics, Pakistan Institute of Engineering and Applied Sciences, P.O. Nilore, Islambad 45650, Pakistan}
\author{Can Wang}
\affiliation{Beijing National Laboratory for Condensed Matter Physics, Institute of Physics, Chinese Academy of Sciences, Beijing 100190, China}
\affiliation{CAS Center for Excellence in Topological Quantum Computation and School of Physical Sciences, University of Chinese Academy of Sciences, Beijing 100049, China}
\affiliation{Songshan Lake Materials Laboratory, Dongguan, Guangdong 523808, China}

\author{Xiulai Xu}
\email{xlxu@iphy.ac.cn}
\affiliation{Beijing National Laboratory for Condensed Matter Physics, Institute of Physics, Chinese Academy of Sciences, Beijing 100190, China}
\affiliation{CAS Center for Excellence in Topological Quantum Computation and School of Physical Sciences, University of Chinese Academy of Sciences, Beijing 100049, China}
\affiliation{Songshan Lake Materials Laboratory, Dongguan, Guangdong 523808, China}


\begin{abstract}
   Monolayer transition metal dichalcogenides have recently attracted great interests because the quantum dots embedded in monolayer can serve as optically active single photon emitters. Here, we provide an interpretation of the recombination mechanisms of these quantum emitters through polarization-resolved and magneto-optical spectroscopy at low temperature. Three types of defect-related quantum emitters in monolayer tungsten diselenide (WSe$_2$) are observed, with different exciton g factors of 2.02, 9.36 and unobservable Zeeman shift, respectively. The various magnetic response of the spatially localized excitons strongly indicate that the radiative recombination stems from the different transitions between defect-induced energy levels, valance and conduction bands. Furthermore, the different g factors and zero-field splittings of the three types of emitters strongly show that quantum dots embedded in monolayer have various types of confining potentials for localized excitons, resulting in electron-hole exchange interaction with a range of values in the presence of anisotropy. Our work further sheds light on the recombination mechanisms of defect-related quantum emitters and paves a way toward understanding the role of defects in single photon emitters in atomically thin semiconductors.

\end{abstract}


\maketitle
\section*{Introduction}
Transition metal dichalcogenides (TMD) monolayers have raised great attentions due to their strong spin-orbit coupling, large exciton binding energy and direct band gap in the visible region \cite{PhysRevLett.105.136805, doi:10.1021/nl903868w,PhysRevLett.108.196802, doi:10.1021/nn305275h, doi:10.1021/nn5021538}. Recently, it has been discovered that a single-layer TMD can be used as a host material for single quantum emitters \cite{ISI:000355620000011,ISI:000355620000008,ISI:000355620000009,ISI:000355620000010,ISI:000354867300013,toth2019single,
doi:10.1021/acs.nanolett.6b04889} at low temperature, which provides a new platform to develop on-chip integrated single photon source and quantum information processing. In addition, the single photon emission has been realized with several approaches, such as heterostructures driven electrically \cite{ISI:000385446300001}, nanoscale strain engineering \cite{doi:10.1021/acs.nanolett.5b03312,ISI:000401906400001,ISI:000401906700001,ISI:000445107900013} and sub-nm focused helium ion irradiation \cite{ISI:000472480700002}, which are mostly defect related. Meanwhile, the properties of such a 2D host of quantum emitters have been intensely investigated, including 3D localized trions in heterostuctures \cite{doi:10.1021/acs.nanolett.7b05409}, manipulation of fine structure splitting (FSS) \cite{PhysRevB.99.045308} and photon-phonon interaction \cite{ISI:000460166500015}. Furthermore, the optical initialization of a single spin-valley in charged WSe$_2$ quantum dots \cite{ISI:000467053100017} and the ability to deterministically load either a single electron or single hole into a Van der Waals heterostructure quantum device via a Coulomb blockade \cite{ISI:000467053100020} have been demonstrated, which enable a new class of quantum-confined spin system to store and process information. However, the origin of the 2D host of quantum emitters is still vague.

Up to now, several theoretical calculations have been performed to investigate defect-mediated charge recombination in TMD monolayers \cite{doi:10.1021/jacs.9b04663,doi:10.1021/acs.nanolett.7b04374,doi:10.1021/acsnano.9b02316,ISI:000439319000050}. Additionally, it has been suggested that vacancies prevailing in monolayer WSe$_2$ may be involved in the single photon emission \cite{PhysRevLett.119.046101}. However, the defect-related recombination mechanisms in 2D host of quantum emitters have not been clearly illustrated experimentally. Especially, the origin of the emissions with remarkably large anomalous exciton g factor and FSS still remains unclear. Magneto-optical spectroscopy can be used to get a deeper insight to the properties of the localized excitons. For example, zero-field splitting is proportional to the spatial overlap of electron and hole wave function and is related to the symmetry of the excitons from which the spatial extension and the symmetry can be deduced \cite{PhysRevB.58.16221}, and the Zeeman splitting can be used to verify the contribution of the spin, valley and orbital magnetic moment in TMD monolayer \cite{PhysRevB.77.235406, PhysRevLett.108.196802,PhysRevLett.114.037401,ISI:000349934700017,ISI:000349934700018,PhysRevLett.113.266804,ISI:000462086600027,ISI:000371020600009}.

In the work, we propose a model to explain the physical picture behind the anomalous magnetic response of single photon emitters in monolayer WSe$_2$ measured by polarization-resolved and magneto-optical spectroscopy. Three types of quantum dots are identified by different exciton g factors and FSSs in monolayer WSe$_2$ which generate single photons. A theory model is proposed to identify the origin of three types of quantum dots. These single quantum emitters come from the different defect-induced transitions, leading to distinct g factors. The excitons originating from different transitions between defect-induced energy levels and conduction band or valance band possess different spatial overlap of the electron and hole wave functions, which result in an anomalous FSS. This work presents an interpretation of the recombination mechanisms experimentally, shedding more light on the origin of these quantum emitters in layered 2D materials.

\section*{Results}
\subsubsection*{\textbf{Photoluminescence spectra at zero magnetic field}}

\begin{figure}
\includegraphics[scale=0.5]{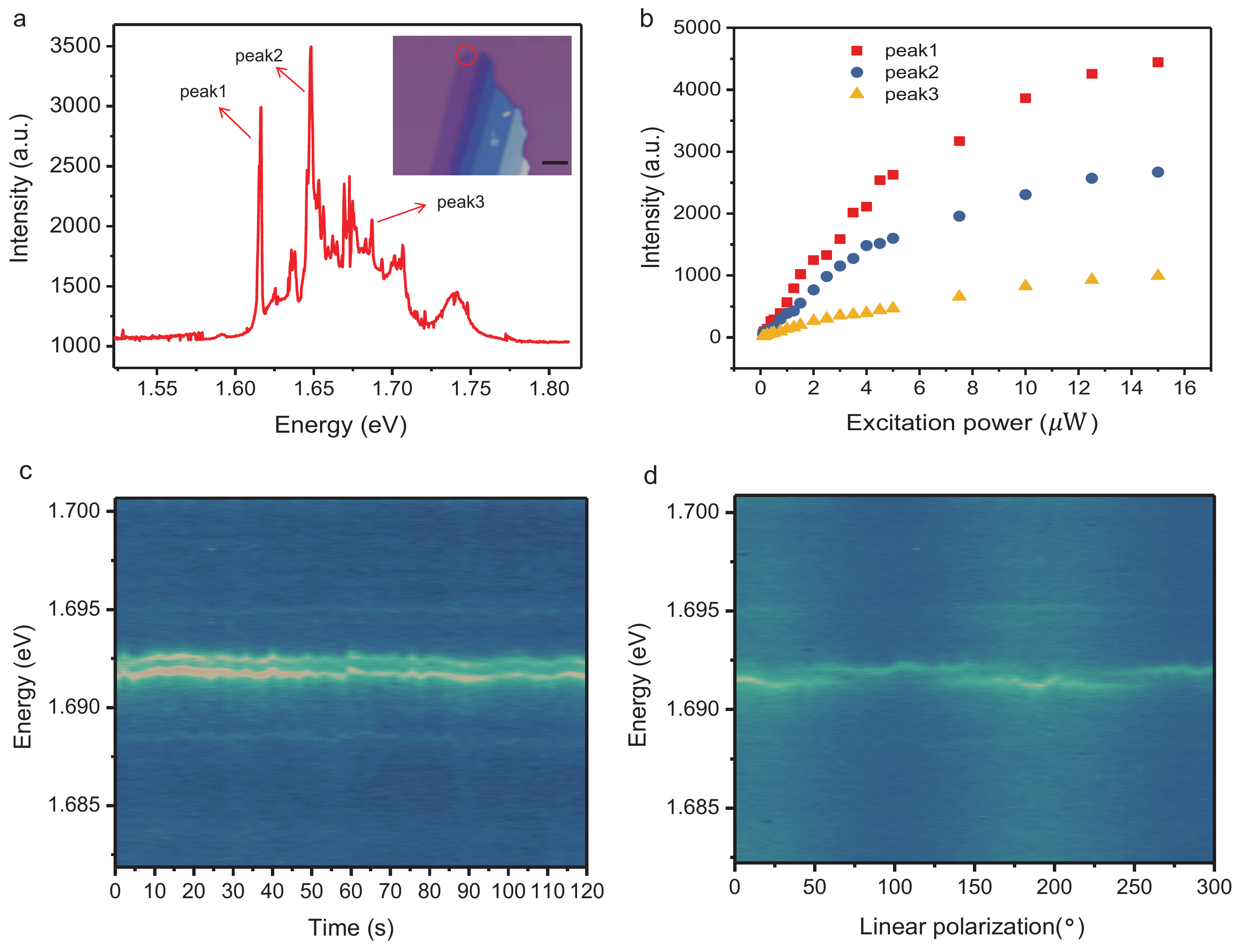}
\caption{\label{f1} Optical properties at zero magnetic field. (a) PL spectrum at the edge of WSe$_2$ flake at 4.2 K with an excitation wavelength of 532 nm and a power of 5 $\mu$W. Inset:  Optical microscopy image of the sample with a scale bar of 5 $\mu$m. The red circle indicates the position of PL measurement. (b) The excitation-power dependence of PL intensity of the sharp peaks in (a). The intensities show saturation behaviour. (c) Time trace of a PL spectrum. The same wandering behaviour of the doublet suggests they originate from same quantum dot. (d) Contour plot of linear-polarization-dependent PL spectra. A pair of cross-linearly polarized spectral doublet is observed.}
\end{figure}

Photoluminescence (PL) spectra at 4.2 K were measured at the edge of monolayer WSe$_2$, as shown in Fig. \ref{f1}(a), in which very sharp peaks emerge. The position of the measurement is indicated with the red circle in the inset. The intensities of the sharp lines are stronger than those of the delocalized neutral valley exciton. With increasing the temperature, the intensities of these peaks decrease until disappearing with linewidth broadening dramatically (see Supplementary Figure 1). Meanwhile, these sharp peaks show saturation behaviours with increasing excitation power, as shown in Fig. \ref{f1}(b). The sharp emission lines originate from the highly spatially localized excitons and usually arise at the edge of the monolayer WSe$_2$. Additionally, the cascaded emission of single photons from the biexciton has been reported in such quantum emitters \cite{ISI:000387349400001} and no obvious Zeeman shift is observed with in-plane magnetic field (Voigt configuration) as shown in Supplementary Figure 3, which are analogous to the property of quantum dots. So these sharp lines can be treated as transitions from quantum-dot like emitters \cite{ISI:000355620000011,ISI:000355620000008,ISI:000355620000009,ISI:000355620000010,ISI:000354867300013,ISI:000367839600009}, which universally result from imperfections in exfoliation or strain effect \cite{doi:10.1021/acs.nanolett.5b03312,ISI:000401906400001,ISI:000401906700001,ISI:000445107900013,ISI:000377507100030,ISI:000374230700022}. Fig. \ref{f1}(c) shows time trace of PL emission from one quantum dot. The  doublet shows spectral wandering with a range of a few hundred $\mu$eV. The synchronized wandering behaviour of the doublet strongly suggests that they originate from the same quantum dot and the splitting is associated with FSS due to the electron-hole exchange interaction. The corresponding linear-polarization-dependent PL results are shown in Fig. \ref{f1}(d). The pair of cross-linearly polarized spectral doublet further demonstrate that the two sharp lines originate from an optically active quantum dot\cite{PhysRevLett.76.3005,PhysRevB.65.195315} embedded in WSe$_2$ monolayer.

\subsubsection*{\textbf{Magneto-optical properties}}

\begin{figure}
\includegraphics[scale=0.5]{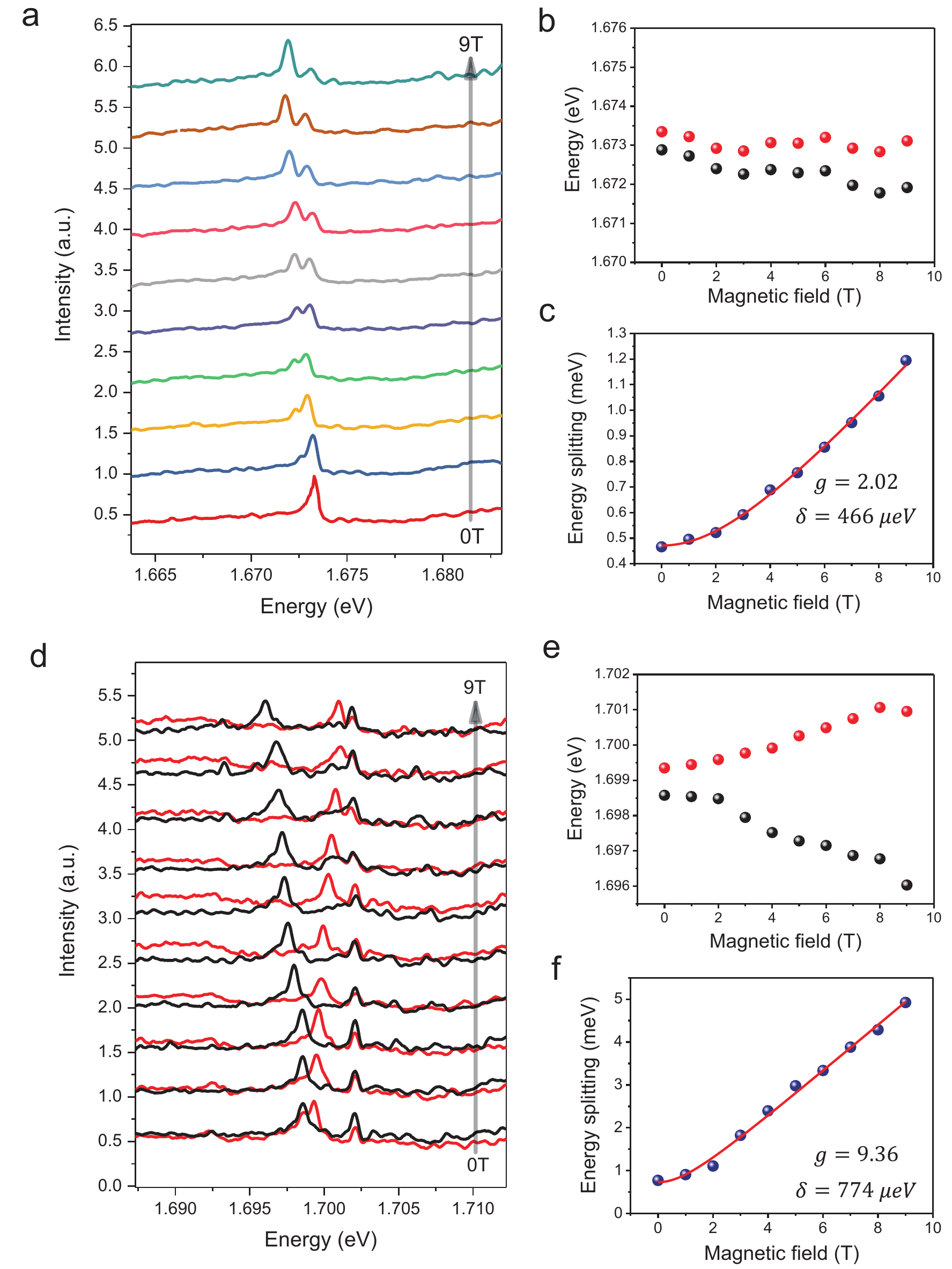}
\caption{\label{f2} Magneto-optical measurement of two types of quantum dots in Faraday configuration. (a) PL spectra of QD1 as a function of the applied magnetic field with a step of 1 T. (b) Fitted peak energy of QD1. (c) Energy splitting of QD1 against magnetic field with an extracted g factor of 2.02 and zero-field splitting of 466 $\mu$eV. (d) Polarization-resolved spectra of QD2 as a function of the applied magnetic field. The single peak at the high energy side of the doublet originates from other quantum emitters. (e) Extracted energies of the $\sigma^+$ (red circles) and $\sigma^-$ (black circles). (f) Energy splitting as a function of the applied magnetic field. The fitted result is represented by the red line, with a g factor of 9.36 and zero-field splitting of 774 $\mu$eV. The spectra in (a) and (d) are normalized for easy comparison.}
\end{figure}

To further clarify the recombination mechanisms of the quantum emitters, we measured the polarization-resolved and magneto-optical PL spectra of many quantum emitters. Three types of quantum dot emissions with different g factors and FSSs were observed. Fig. \ref{f2} depicts the magneto-optical properties of two types of quantum dots, (a)-(c) for one type and (d)-(f) for the other. Fig. \ref{f2}(a) shows the PL spectra as a function of an applied magnetic field perpendicular to the sample (Faraday configuration). The extracted energies of the PL peaks by two-peak Lorentz fitting are shown in Fig. \ref{f2}(b). The two peaks exhibit correlated spectral wandering, suggesting they arise from the same quantum dot. The energy splitting increases with increasing the magnetic field, as shown in Fig. \ref{f2}(c). A hyperbolic dispersion is used to fit the data \cite{ISI:000355620000011,ISI:000355620000008,ISI:000355620000009}
\begin{equation}
 E(B)=\sqrt{(g\mu_BB)^2+\delta^2}\,
\label{wkb}
\end{equation}
where $\mu_B$ is the Bohr magneton, g is the exciton g-factor, and $\delta$ is the zero-field splitting (that is, FSS). The fitted zero-field splitting is about 466 $\mu$eV, which is related to electron-hole exchange interaction. A exciton g factor of 2.02 can be extracted from the fitting, which is similar to that of excitons in self-assembled III-V quantum dots \cite{PhysRevLett.76.3005,PhysRevB.65.195315,RevModPhys.85.79,PhysRevB.70.235337,PhysRevB.85.165323,PhysRevApplied.8.064018}. We label this quantum emitter as QD1. We also measured another type of emitter (labelled QD2) with distinct magnetic response, as shown in Fig. \ref{f2}(d). The single peak at the high energy side of doublet originates from another quantum dot. As the splitting of the doublet increases with magnetic field, the polarization of doublet changes from linear to circular. The extracted energies of the $\sigma^+$ and $\sigma^-$ detection are shown in Fig. \ref{f2}(e) and energy splitting as function of magnetic field is plotted in Fig. \ref{f2}(f). The zero-field splitting and exciton g factor are extracted with values of 774 $\mu$eV and 9.36, respectively. These values are in consistence with previous work on quantum dots in WSe$_2$, in which the g factors range from 6 to 12 \cite{ISI:000355620000011,ISI:000355620000008,ISI:000355620000009,ISI:000355620000010,ISI:000354867300013}.

\begin{figure}
\includegraphics[scale=0.5]{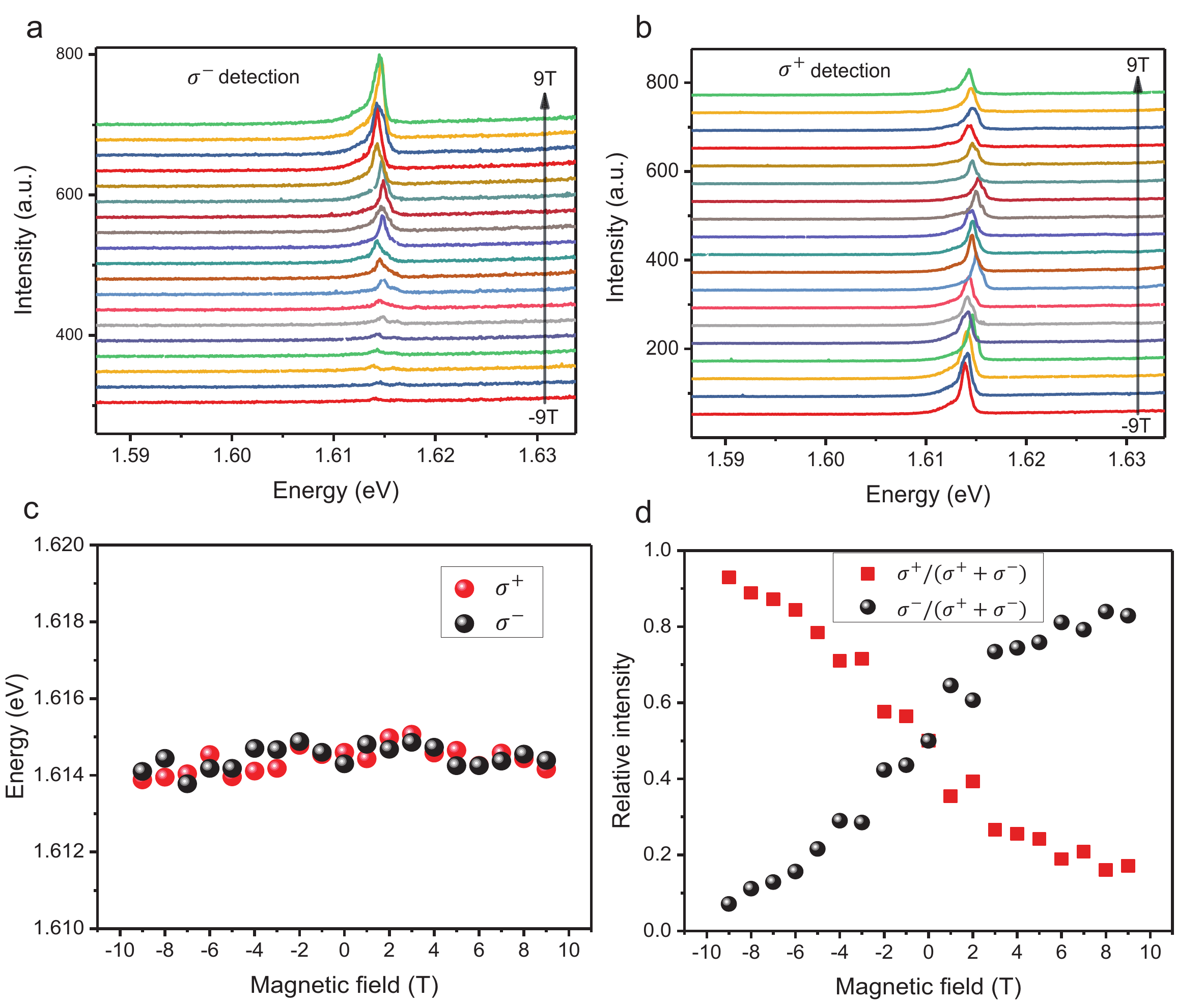}
\caption{\label{f3} Polarization-resolved and magneto-optical spectroscopy of QD3 in Faraday configuration. PL intensity of $\sigma^-$ (a) and $\sigma^+$ (b) component as a function of the applied magnetic field. (c) Extracted central energies of the $\sigma^+$ (red) and $\sigma^-$ (black) detection associated with QD3 against magnetic field, with no observable splitting and Zeeman shift. (d) Extracted relative intensity of the $\sigma^+$ (red squares) and $\sigma^-$ (black circles) components as a function of the applied magnetic field. The range of magnetic field is from -9 to +9 T with a step of 1 T. }
\end{figure}

Interestingly, we also observed another type of emitter (labelled QD3). Fig. \ref{f3}(a) and (b) show PL spectra resolved in circular polarizations $\sigma^+$ and $\sigma^-$, respectively. The extracted peak energies of the $\sigma^+$ and $\sigma^-$ detection are shown in Fig. \ref{f3}(c). QD3 exhibits a single peak with unobservable splitting and unobservable Zeeman shift compared with QD1 and QD2 discussed above, which is within the resolution of our experiment and the spectral wandering for this emitter. Fig. \ref{f3}(d) shows peak intensities of the $\sigma^+$ and  $\sigma^-$ detection as a function of the magnetic field. The intensity of the $\sigma^-$ increases with increasing the magnetic field while the $\sigma^+$ shows completely opposite behaviour, which is similar to the behaviour of intrinsic valley excitons in TMD monolayer \cite{ISI:000414649300008,ISI:000444494800005}. Besides the examples in the main text, more emitters with similar properties to these types are shown in Supplementary information.

\section*{Discussion}
\begin{figure}
\includegraphics[scale=0.65]{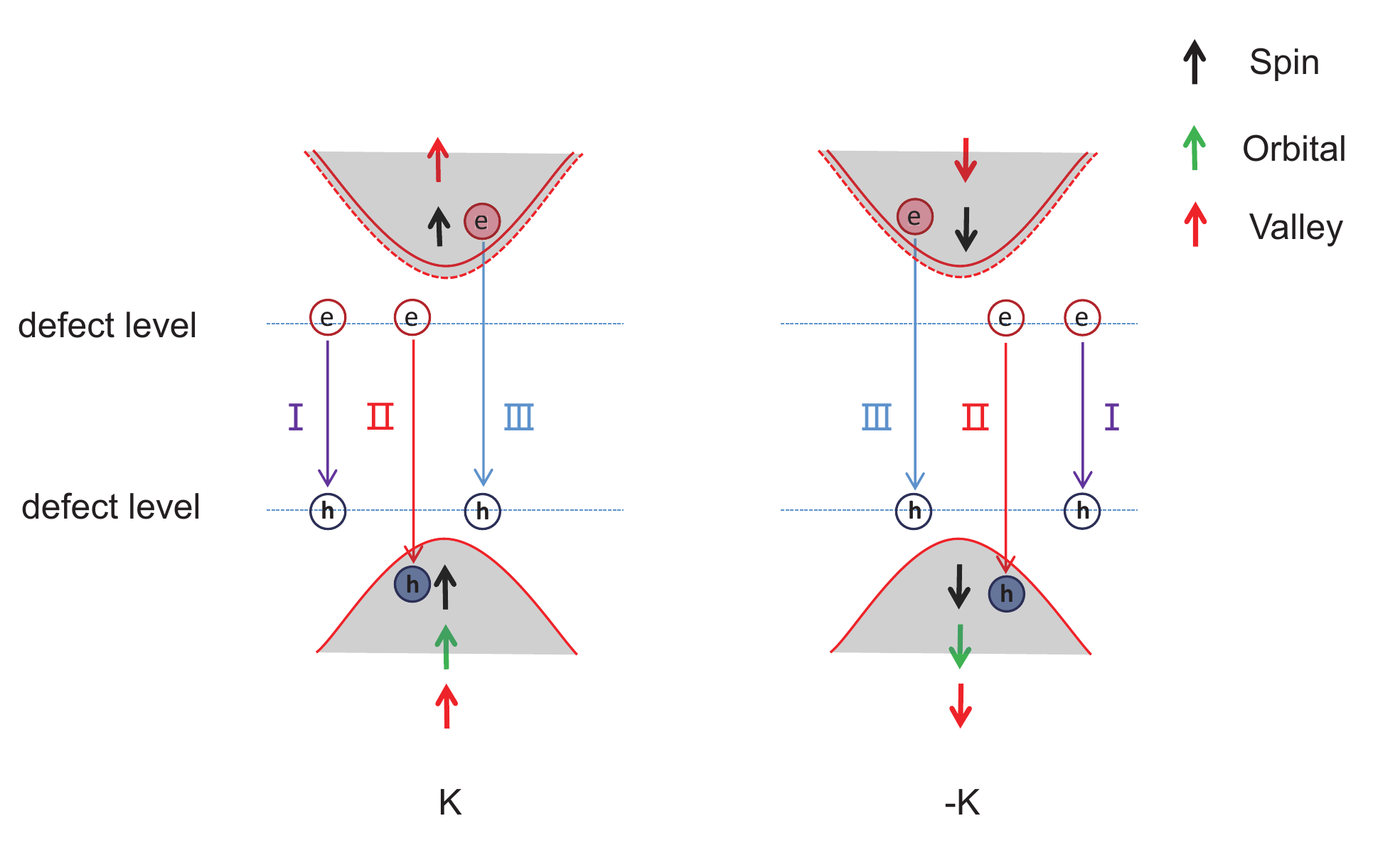}
\caption{\label{f4}  Schematic diagram of recombination mechanisms assisted by defect levels. Type \uppercase\expandafter{\romannumeral1} represents transitions between two defect levels. Type \uppercase\expandafter{\romannumeral2} (\uppercase\expandafter{\romannumeral3}) represents the transitions from defect level (conduction band) to valence band (defect level). Type \uppercase\expandafter{\romannumeral1}, \uppercase\expandafter{\romannumeral2} and \uppercase\expandafter{\romannumeral3} correspond to QD1, QD2 and QD3, respectively. A small spin-orbit splitting arises in the conduction band and the dashed line in the conduction band depicts spin-forbidden state for the ground state of WSe$_2$. The spin, orbital and valley magnetic moment are represented by short black, green and red arrows.  Open and solid circles refer to trapped and non-trapped states, respectively. }
\end{figure}

Generally, the total magnetic moment of intrinsic valley excitons in TMD monolayer consists of three parts: spin, orbital and valley magnetic moment \cite{PhysRevB.77.235406, PhysRevLett.108.196802,PhysRevLett.114.037401,ISI:000349934700017,ISI:000349934700018,PhysRevLett.113.266804,ISI:000462086600027,ISI:000371020600009}, as shown in Fig. \ref{f4}. The spin does not contribute to Zeeman splitting because the optical valley selection rule requires that the valence and conduction band have the same spin. And the valley magnetic moment associated with the Berry curvature also does not contribute to Zeeman splitting in TMD monolayer \cite{PhysRevLett.114.037401,ISI:000371020600009}. However, the valence and conduction bands are composed of completely different transition metal atomic orbital. The conduction band is constituted by $d_{z^2}$ orbitals, corresponding to an azimuthal orbital angular momentum with $l_z=0$. While the valence bands is constituted by the hybridized orbitals $d_{x^2-y^2}\pm id_{xy} $ that have orbital angular momentum $l_z=\pm 2h$ in the K and -K valleys, respectively. Therefore in Faraday configuration, the term of orbital magnetic moment will generate a Zeeman splitting of 4 $\mu_B$B. Considering these three parts, the intrinsic valley exciton possesses a g factor of 4.

Compared to the intrinsic valley exciton, these quantum dots embedded in WSe$_2$ monolayer have distinct magneto-optical properties, including exciton g factor and FSS, which usually result from point defects or strain effects. It is worth noting that these quantum emitters in our sample mainly result from defects other than strain effect since no intentional strain is introduced. And the emissions are assumed to be neutral excitons which usually appear without applied electric field \cite{ISI:000467053100017,ISI:000467053100020}. The defects introduce various trapping energy levels \cite{PhysRevLett.119.046101, PhysRevLett.121.057403, doi:10.1021/acs.nanolett.7b04374, doi:10.1021/jacs.7b02121, doi:10.1021/jacs.8b13392, doi:10.1021/jacs.9b04663} within the electronic band gap of the WSe$_2$, thus providing possibilities for various transitions. For different transitions, distinct magnetic response can be predicted.

Similar to the case of intrinsic valley excitons, the total magnetic moment of these localized excitons can be considered from the three terms: spin, orbital and valley magnetic moment. The valley contribution to Zeeman splitting for each type is expected to be zero since defect states, valance and conduction band should experience the same shift with valley under magnetic field. For type \uppercase\expandafter{\romannumeral2} and type \uppercase\expandafter{\romannumeral3} shown in Fig. \ref{f4}, the transitions between bands and defect levels, the spin contribution is expected to be zero according to the optical selection rules, in consistent with the intrinsic valley excitons. The difference between them mainly results from the orbital contributions. When the holes or electrons are trapped in the defect, the orbital magnetic moment will decrease as the confinement increases \cite{PhysRevB.85.165323}. The orbital contribution of hole trapped in defect level near valence band is smaller than that of electron trapped in defect level near the conduction band due to a more localized wave function \cite{doi:10.1021/jacs.9b04663}, which can be ignored. For type \uppercase\expandafter{\romannumeral2}, which represents the transition from defect level to valence band, the total orbital magnetic moment contributions include the orbital contribution of valence band of 4 $\mu_B$ and the orbital contribution of electron in defect level, therefore leading to a relatively large and widespread exciton g factor. This type of emitter is observed in QD2 and has been reported in other works, in which the exciton g factors range from 6 to 12 \cite{ISI:000355620000011,ISI:000355620000008,ISI:000355620000009,ISI:000355620000010,ISI:000354867300013}. While for type \uppercase\expandafter{\romannumeral3}, the contributions of orbital magnetic moment include the orbital contribution of conduction band of 0 and the orbital contribution of hole in defect level which can be ignored. It will lead to unobservable splitting under magnetic field. This type of emitter is more frequently observed in our samples (QD3) because usually W vacancies are the dominant point defects in monolayer WSe$_2$ \cite{PhysRevLett.119.046101}, which will introduce defect states near the valance band \cite{doi:10.1021/jacs.9b04663,ISI:000439319000050}. For type \uppercase\expandafter{\romannumeral1} transition, the electron and hole are both trapped by the defect-induced levels, which are localized around the defect, similar to semiconductor quantum dots \cite{PhysRevB.65.195315,PhysRevLett.82.1748,PhysRevB.70.235337,PhysRevB.85.165323}. A moderate g factor and FSS can be predicted, which is consistent with the property of QD1.

Meanwhile, for QD2, the electron is confined in the defect, causing a smaller exciton Bohr radius, which enhances the spatial overlap between the electron and hole wave function, resulting in strong electron-hole exchange interaction \cite{PhysRevB.70.235337,PhysRevLett.82.1748}. Therefore a large FSS can be observed. However, for QD3, the electron is in conduction band that indicates a large spatial extension of electron, the overlap between electron and hole wave function dramatically decreases compared to QD2. So the electron-hole exchange interaction is strongly suppressed, resulting in an unobservable FSS.

The intensity of QD3 exhibits different magnetic field dependence with different helicity, as shown in Fig. \ref{f3}, which results from the thermal equilibrium-related valley property \cite{ISI:000414649300008,ISI:000444494800005}. Under positive magnetic field, all bands will experience a large upshift (downshift) in the K (-K) valley (as shown in Supplementary Figure 5). The electrons in different valleys will have energy difference, resulting in unequal distributions in the two valleys following the case near thermal equilibrium. It means that the electrons tend to occupy the conduction band in -K valley. And the case for negative magnetic field is opposite. The unequal distributions in the two valleys lead to the magnetic field dependence of intensity with different helicity, as shown in Fig. \ref{f3}.

In summary, we have identified three types of quantum emitters in TMD monolayer with polarization-resolved and magneto-optical microscopy. A theory model is proposed to interpret the recombination mechanisms of defect-related quantum emitters. The various transitions between the defect-induced levels, valance and conduction band lead to anomalous exciton g factor. Meanwhile, the spatial overlap with a range of values between electron and hole wave function lead to various FSS. Our work provides a theory model to explain the origin of anomalous large g factor and FSS, shedding more light on the origin of these quantum emitters in layered 2D materials, furthermore paving a way toward understanding the role of quantum dots in single photon emitters in atomically thin semiconductors and providing an unique platform for solid state quantum information processing \cite{PhysRevB.88.075404,PhysRevX.4.011034}.

\section*{Methods}
\subsubsection*{\textbf{Sample fabrication}}
Single layer WSe$_2$ was obtained by mechanical exfoliation by Scotch tape from a synthetically grown single crystal, and transferred onto a Si substrate with 300 nm SiO$_2$ on top.

\subsubsection*{\textbf{Photoluminescence spectroscopy}}
The sample was placed on three-dimensional piezoelectric stage in a helium gas exchange cryostat equipped with superconducting magnet, which can supply vertical magnetic field from -9 T to 9 T and horizontal magnetic field from -4T to 4T. The sample was cooled down to about 4.2 K and the temperature was changed with a heater. Photoluminescence spectroscopy was performed using a confocal microscope setup. The sample was non-resonantly excited by 532-nm continuous laser. And the emission was collected using a x50 objective with a numerical aperture of 0.55 and directed to a grating spectrometers through optical fiber where a 1200 $gmm^{-1}$ gratings was used for high-resolution spectra. A liquid-nitrogen-cooled charge-coupled-device camera was used to detect the PL signal. The polarization of the emission was measured by using a $\lambda/2$ plate or $\lambda/4$ plate followed by a polarizer.

\section*{Data Availability}
The authors confirm that the data supporting the findings of this study are available within the article. Related additional data are available on reasonable request from the authors.

\section*{Acknowledgements}

This work was supported by the National Natural Science Foundation of China under Grants No. 51761145104, No. 11934019, No. 61675228, No. 11721404 and No. 11874419; the Strategic Priority Research Program (Grant No. XDB28000000), the Instrument Developing Project (Grant No. YJKYYQ20180036), the Interdisciplinary Innovation Team of the Chinese Academy of Sciences, and the Key RD Program of Guangdong Province (Grant No.2018B030329001).

\section*{Author Contributions}

X. Xu, C. Wang and M.Rafiq conceived and planned the project; J. Dang prepared the sample; J. Dang, S. Sun, X. Xie, Y. Yu, K. Peng, C. Qian, S. Wu, F. Song, J. Yang, S. Xiao, L. Yang, Y. Wang and X. Xu performed optical measurement; J. Dang, S. Sun, X. Xie, Y. Yu and X. Xu analysed the data. J. Dang, X. Xie, and X. Xu wrote the manuscript with contributions from all others.

\section*{Competing Interests}

The authors declare no competing interests.

\section*{References}
\providecommand{\newblock}{}


\section*{Additional Information}

\textbf{Supplementary information} is available for this paper at.

\textbf{Correspondence }should be address to X. Xu.



\end{document}